\documentstyle[aps,pre,epsf]{revtex} 

\title{Spectral Determinant Method for Interacting N-body Systems 
Including Impurities}

\author{Z.\ \'A.\ N{\'e}meth, J. Cserti and G.\ Vattay}

\address{E{\"o}tv{\"o}s University, Department of Physics of Complex Systems, 
H-1117 Budapest, P\'azm\'any P{\'e}ter s{\'e}t\'any 1/A, Hungary}

\begin{document} 

\draft
     

\maketitle

\begin{abstract}
A general expression for the Green's function of a system 
of $N$ particles (bosons/fermions)  interacting by contact
potentials, including impurities with Dirac-delta type
potentials is derived.
In one dimension for $N>2$ bosons from our 
{\it spectral determinant method\/} the numerically 
calculated energy levels  agree very well with those obtained from 
the exact Bethe ansatz solutions while they are an order of magnitude more 
accurate than those found by direct diagonalization. For $N=2$ bosons
the agreement is shown analytically.  
In the case of $N=2$ interacting bosons and  one
impurity, the energy levels are calculated numerically 
from the spectral determinant of the system. The spectral determinant 
method is applied to an interacting fermion system including an impurity to
calculate the persistent current at the presence of magnetic field. 
\end{abstract}

\pacs{PACS numbers: 71.10.-w,  71.15.-m, 71.55.-i}

\section{Introduction} \label{sec:intro}
Due to the remarkable progress in the understanding of many features of 
mesoscopic physics the role of electron-electron interactions has been
actively studied in such systems. 
Two frequently studied problems are the 
interaction induced delocalization of two particles in disordered systems 
(we shall use the abbrevation TIP  as it is common in the literature 
for two interacting particles)\cite{Dorokhov,Shepelyansky} 
and the persistent current (PC)
first predicted by B\"uttiker et al.\ \cite{Buttiker1} and observed 
in metallic rings by Levy et al.\ \cite{Levy} and Chandrasekhar 
et al.\ \cite{Chandrasekhar}. 
Commonly used numerical approaches to these problems are the Green's
function (GF) method\cite{Oppen1,Song1,Ortuno}, the  
diagonalization method\cite{Pichard1,Berkovits1,Berkovits2}, 
the decimation method\cite{Leadbeater} and density-matrix
renormalization-group (DMRG) method\cite{Chandross}.  
The many-body systems including impurities are 
frequently modeled by the Anderson-Hubbard tight-binding Hamiltonian 
\cite{Ortuno} given by interacting electrons (with on-site interaction), 
and random potentials (with random site energy) of the impurities on a
lattice. 
In the TIP problem the GF of the full Hamiltonian is 
obtained from  Dyson's equation considering the interaction 
Hamiltonian as a perturbation. The two-particle 
localization length is then defined via a particular trace of the GF. 
In the PC problem  the interplay between the
interactions and impurities has generated much interest 
and several theoretical methods are developed (see e.g.\ Ref.\
\cite{Eckle} and references therein). 
To find the persistent current one needs to calculate the flux
dependence of the ground state energy which can be obtained 
from the lowest pole of the GF as a possible approach.
It is therefore desirable to construct the GF of the system
which includes both impurities and the electron-electron interaction 
in order to treat the TIP and PC problems in a common framework. 

In this paper we develop a formalism to determine the GF of 
N-particle systems in which both the potential of the impurities and the
interaction between the particles are given by contact potentials, 
i.e.\ by Dirac delta functions. 
Unlike in the case of the Anderson tight-binding Hamiltonian, 
in our calculation the positions of the electrons are not restricted 
to lattice sites but can be continuous coordinates. 
In our general formalism both the impurities and the interactions 
are treated as perturbations where the interaction or impurity
strengths are the small parameters,
 and the GF is obtained from Dyson's equation. 
The resulting series for the GF can be expressed in terms of the GFs of 
the noninteracting many particle systems without impurities. 
If  both the impurities and the interaction are given by contact
potentials, the series can be given in a closed form, namely as 
the ratio of two determinants including only the GF of the unperturbed
system. A similar (i.e.\ ratio of two determinants) 
form of the GF was found by Grosche\cite{Grosche} 
for a system containing only {\it one\/} electron interacting with  
impurities. We have used this form of the GF in Ref.\  \cite{Chaos-cikk} 
to investigate the diffraction of the electron by the impurities, 
and in Ref.\ \cite{Crossover} to study the crossover 
behavior of the localization problem in a waveguide with 
periodically placed identical point-like impurities.
In this paper  an extension of Grosche's approach to treat 
many-body interacting systems including impurities is presented.  
In our formalism we consider both the spinless bosons and
the spin-full fermions.   
The many-body  wave functions 
are the symmetrized products of the one-particle eigenstates of the system. 
In our calculations the bosonic/fermionic feature of the many-body
system is fully incorporated. 
The energy eigenvalues of many-body systems are given 
by the poles of the GF. It implies that the energy levels are 
the zeros of the determinant in the denominator of the GF. 
 We claim that the energy levels determined through this spectral 
determinant are much more accurate than those found by the traditional
diagonalization methods. Therefore, here we propose the method of
spectral determinants (SD) as a new approach to this problem.  
As it is seen below the indices of the matrix elements of the determinants 
are continuous variables. To evaluate the necessary SD in this
case, one can use any complete orthogonal set as a basis. 
In our examples below this will be demonstrated using one-particle 
eigenstates. 
   
As an application and to test our method, we consider the one-dimensional
N-body problem in which the bosonic particles interact via a Dirac delta
potential. It is well known that the exact result for the energy eigenvalues
can be obtained from  the Bethe ansatz
\cite{Bethe,Lieb-I,Lieb-II,Yang-Yang,Mattis},
so it is possible to compare our results with the exact ones. 
It is analytically proved that for the case of two interacting bosons the 
energy eigenvalues obtained from our method are {\it identical\/} 
to those found from the Bethe ansatz solutions. 
Moreover, in the case of three and four particles our 
numerical results agree very well with the Bethe ansatz solutions. 
Our results are compared with those obtained from direct diagonalization of
the Hamiltonian, and it is demonstrated that in general our results are 
more than ten times more accurate. 
For higher energies it is shown that the direct 
diagonalization method is very inaccurate, which seriously limitates its
applicability in numerical calculations.    

To see how much more effective our method is, we shall consider two  
nontrivial examples as a further application.  We present our results
for the system of two interacting bosons and one impurity, and for 
three interacting spin-full fermions and one impurity. 
In these cases no Bethe ansatz type of solution is known. 
Increasing the number of one-particle wave functions the energy
eigenvalues converge.  This way all the energy levels are
calculated to four significant figures, and are taken as the exact
energy levels. The errors of the energy eigenvalues found from our SD 
method and from the direct diagonalization of the Hamiltonian are
compared using the same number of one-particle wave functions. 
In general, our method gives again an order of magnitude more accurate results 
than the diagonalization of the Hamiltonian.
The accurate energy levels are necessary for calculating e.g.\ the
persistent current which is the derivative of the ground state energy
with respect to the applied magnetic flux enclosed by the system. 
As a demonstration we present our calculation of the persistent current
for the interacting three-particle fermion system.

Since in the above mentioned TIP and PC problems the Anderson-Hubbard
tight-binding Hamiltonian\cite{Ortuno} 
can be regarded as a discretized version of the
continuous model with Dirac delta potentials for both the interaction among
the particles and with the impurities, we expect that our SD
method might be applied successfully in numerical simulations. 
The work along this line is in progress.       

The rest of the paper is organized as follows. 
In section \ref{int+impuri} we give our most general form of the GF 
for interacting N-particles systems including a finite number of
impurities. In section \ref{spin-fermion} the GF is given for
spin-full case.  In all these sections our results for the GF 
are given as the ratio of two determinants expressed in terms of the 
GF of the noninteracting N-particle system. 
In section \ref{test} our method is tested for
interacting bosons for which the exact Bethe ansatz solutions are known. 
Our results are compared with the exact ones obtained from the 
Bethe ansatz solution and with those found from the numerical 
diagonalization of the corresponding Hamiltonian.   
In section \ref{2p+impuri} we calculate the energy levels 
for a system of two interacting bosons plus one impurity and
compare the results with those obtained from the direct diagonalization of 
the Hamiltonian.
In section \ref{3pfermion+impuri} the calculation of the persistent
current is presented for three interacting fermions and one impurity.   
Our conclusions are given in section \ref{veg}.
In Appendix \ref{app-1} the derivation of the GF is given for the case of two 
interacting particles and $M$ impurities. 
In Appendix \ref{2boson} we show that from our method the exact Bethe 
ansatz solution can be obtained analytically for two interacting bosons.

\section{Many-body boson systems with interaction and impurities}
\label{int+impuri}

In this section we consider the most general case, namely the system of
$N$ interacting spinless bosons plus $M$ impurities with different 
strengths of potentials. 
To highlight the method to be applied it is useful to discuss the  
simple system of two noninteracting bosons with one impurity.
It turns out that the full GF is a ratio of two structured
determinants. However, both determinants are so called 'continuous
matrices' i.e.\  their indices are continuous variables.   
This feature of the matrices arises from the many body character of
the system. 

The GF of the Hamiltonian $H=H_0+H_1$ is given by 
\begin{equation}
G = {\left(E - H_0 - H_1 \right)}^{-1}.
\label{Green-de}
\end{equation}
The Hamiltonian $H_0$ of  $N$ identical noninteracting particles 
can be written as
\begin{equation}
H_0({\bf x})=\sum_{i=1}^N h(x_i),
\label{H0-op}
\end{equation} 
where the position of the $i$th particle is denoted by 
$x_i$, and ${\bf x}= (x_1,x_2,\dots,x_N)$. 
In general, the $x_i$ are  $d$ dimensional vectors.
In the examples presented below only one-dimensional systems 
are studied. 
The one-particle Hamiltonian $h(x)$ is given by
\begin{equation}
h(x)= -\nabla^2+V(x), 
\label{free-H}
\end{equation}
where $V(x)$ is the potential in the one-particle problem. 
Hereafter we shall use  $\hbar = 2 m = 1$ units, where $m$ is the mass of
the particles.
We assume that both the eigenstates and the energy eigenvalues are 
known for the Hamiltonian $h$ of the one-particle problem.
In many applications the potential $V(x)$ is taken to be zero inside 
the box and infinity at its boundaries. 
The other typical case is when the potential
$V(x)$ is zero inside the box and periodic boundary conditions are
applied. In this paper the latter is used in our examples. 

The Hamiltonian $H_1$ represents the effect of the impurity, and can be
written as 
\begin{equation}
H_1({\bf x})=\kappa \sum \limits_{p=1}^N \delta(x_p-u),
\label{szenny}
\end{equation}
where the position of the impurity is denoted by $u$, while $\kappa$ is 
the strength of the Dirac delta potential. 
The full GF of
$H_0+H_1$ satisfies Dyson's equation 
\begin{equation}
G = G_0 + G_0 H_1 G,
\label{Dyson}
\end{equation}
where $G_0$ is the GF of the Hamiltonian $H_0$.

To proceed further, it is instructive to consider only two noninteracting
particles, i.e.\ $N=2$. The result can be easily generalized then to 
the case $N > 2$. 
For two noninteracting particles the GF in coordinate representation is given
\cite{Economou} by
\begin{equation}
G_0(x_1,x_2|x_1',x_2') =\sum \limits_n {\psi_n(x_1,x_2)
\psi_n^*(x_1',x_2')
\over E-E_n },
\label{2p-G0}
\end{equation}
where $\psi_n(x_1,x_2)$ is the $n$th eigenstate of the two noninteracting 
particles, and the corresponding eigenvalue is $E_n$. 
In case of bosons the wave function is symmetric for the permutation 
of its variables, i.e.\ $\psi_n(x_1,x_2) = \psi_n(x_2,x_1)$.   

The iterative solution of Dyson's equation results in the 
series
\begin{eqnarray}
\lefteqn{G(x_1,x_2|x_1',x_2') =  G_0(x_1,x_2|x_1',x_2')+
\int G_0(x_1,x_2|y_1,y_2) \ H_1(y_1,y_2) \ G_0(y_1,y_2|x_1',x_2') 
\ d{y_1}d{y_2}
} 
\nonumber \\ 
& & + \int G_0(x_1,x_2|y_1,y_2) \ H_1(y_1,y_2) 
\ G_0(y_1,y_2|y_3,y_4) \ H_1(y_3,y_4) \ G_0(y_3,y_4|x_1',x_2') \
 d{y_1} d{y_2} d{y_3} d{y_4} + \cdots  . 
\label{dyson-sor}
\end{eqnarray}

After inserting $H_1$ given in Eq.\ (\ref{szenny}) 
into Eq.\ (\ref{dyson-sor}), 
one can perform some of the integrals in the series due to the presence of 
Dirac delta potentials. Finally, making use of the fact that for bosons 
\begin{equation}
G(x_1,x_2|x_1',x_2')=G(x_2,x_1|x_2',x_1')= G(x_2,x_1|x_1',x_2')
= G(x_1,x_2|x_2',x_1'),
\label{permut-2} 
\end{equation}
the series in Eq.\ (\ref{dyson-sor}) leads to
\begin{eqnarray} 
\lefteqn{ G(x_1,x_2|x_1',x_2') = G_0(x_1,x_2|x_1',x_2) + 
2\kappa\int G_0(x_1,x_ 2|y,u) \ G_0(y,u|x_1',x_2') \ dy }
\nonumber \\
&& + (2\kappa)^2 \int G_0(x_1,x_2|y_1,u) \ 
G_0(y_1,u|y_2,u) \ G_0(y_2,u|x_1',x_2') 
\ d{y_1} d{y_2} + \cdots .
\label{sor-1}
\end{eqnarray} 
This is a Neumann series\cite{Arfken}, known from the theory of
integral equations. It can
be rewritten in the more compact form 
\begin{equation}
G(x_1,x_2|x_1',x_2')= G_0(x_1,x_2|x_1',x_2')- \int G_0(x_1,x_2|y_1,u) \ 
K^{-1}(y_1|y_2) \ G_0(y_2,u|x_1',x_2') \ d{y_1} d{y_2},
\label{K-1-op}
\end{equation}
where the operator $K(y_1|y_2)$ is given by 
 \begin{equation}
K(y_1|y_2) = -{1\over 2 \kappa} \ \delta(y_1-y_2) + G_0(y_1,u|y_2,u).
\label{K-def}
\end{equation}
For calculating the inverse of $K$ one can choose a suitable complete
orthogonal set to
represent the operator $K$ as a matrix.   
Using the identity for hypermatrices  
\begin{equation}
{\rm det} \left(
\begin{array}{cc} 
a & b \\
c & d 
\end{array}
\right ) = {\rm det} (d)\  {\rm det}
\left(a - b d^{-1}c \right )
\label{hyperm}
\end{equation} 
(which holds for arbitrary square matrices $a$ and $d$, 
with $\det d \neq 0$), the full GF can be written as the ratio of 
two determinants
 \begin{equation}
G(x_1,x_2|x_1',x_2') = {\det \left[\matrix{G_0(x_1,x_2|x_1',x_2')&
G_0(x_1,x_2|y_2,u)  \cr G_0(y_1,u|x_1',x_2') & K(y_1|y_2)  \cr}\right]\over 
\det K(y_1|y_2)}, 
\label{d/m}
\end{equation}
where the matrix in the numerator is discrete in its first row and
column, but the rest is indexed by the continuous variables
$y_1, y_2 $. The `continuous matrix' $K$ can be given 
in matrix representation. Examples for treating this type of matrices  
will be given in Secs.\ \ref{test}, \ref{2p+impuri} and 
\ref{3pfermion+impuri}. 
Rewriting equation (\ref{K-1-op}) in the form given by Eq.\ (\ref{d/m})
is formal since in the numerator of Eq.\ (\ref{d/m}) the dimension of the 
$1,1$ matrix element of the determinant is unity. However, later it turns out 
that in the more general case in which more than two particles are
included as well as many impurities, the GF can be given similarly 
as a ratio of two determinants.     

Turning to the more general case of $N>2$, the necessary steps to obtain
the full GF are similar to those made in case of $N=2$. The crucial point 
is to make use of the identity of the bosonic GF  
\begin{equation}
G({\bf x}|{\bf x'})=G({\bf Px}|{\bf {P'}x'}),
\label{permutn}
\end{equation}
where ${\bf Px}$ is an arbitrary permutation of the set 
$(x_1,x_2,...,x_N)$. One then finds
\begin{equation}
G({\bf x}|{\bf x'}) = {\det \left[ \matrix{G_0({\bf x}|{\bf x'}) &
G_0({\bf x}|u,{\bf y'}) \cr G_0(u,{\bf y}|{\bf x'}) & K({\bf y}|{\bf y'})}
\right] \over \det K({\bf y}|{\bf y'})}, 
\label{GF-1}
\end{equation}
where 
\begin{equation}
K({\bf y}|{\bf y'}) = -{1 \over N \kappa } \ \delta^{(N-1)} ({\bf
y}-{\bf y'}) + G_0(u,{\bf y}|u,{\bf y'}),
\label{Kmatrix}
\end{equation}
and ${\bf y}=(y_1,y_2,...,y_{N-1})$.
The structure of the GF is similar to that given in Eq.\ (\ref{d/m})
except that now the matrix $K$ depends on some multidimensional variables.
The indices of the `matrix' $K$ in Eq. (\ref{Kmatrix}) are
`continuous' variables. The appearance of this kind of infinite-sized 
matrix is a consequence of the many-body feature of the Hamiltonian. 

We now turn to the case of interacting bosons. The previously
presented method is suitable to handle the more complicated systems in
which the interactions between particles are also taken into account.
The derivation of the full GF of the two interacting bosons with many
impurities is very similar to that
shown before for the case of noninteracting particles,  and the
details are given in Appendix \ref{app-1}. 

In the most general case the Hamiltonian $H_0$ of the 
noninteracting N-particle system without
impurities is given by Eq.\ (\ref{H0-op}).
The rest of the total Hamiltonian of the system with $M$ impurities 
and Dirac delta interactions with $N$ particles is 
\begin{equation}
H_1({\bf x}) = \sum_{i=1}^M \kappa_i \sum_{p=1}^N \delta(x_p-u_i) +
\lambda \sum_{{p,q=1}\atop {p<q}}^N \delta(x_p-x_q),
\end{equation}
where $x_p$ are the positions of the bosons $p=1,\cdots, N$, 
$u_i$ and $\kappa_i$ is the position and the strength of the
impurity $i=1,\cdots ,M$, and $\lambda$ is the strength of the
interactions between particles. For simplifying the notations we
shall use the vector ${\bf x} = (x_1,x_2,...,x_N)$.

To find the full GF, a similar procedure given in Appendix \ref{app-1} 
can also be carried out in
this case, and the final result is 
\begin{equation}
G({\bf x}|{\bf x'}) ={\det \left[ \matrix{G_0({\bf x}|{\bf x'}) & 
G_0({\bf x}|y_1',y_1',{\bf \tilde y'}) & 
G_0({\bf x}|u_1,{\bf y'})& \cdots & G_0({\bf x}|u_M,{\bf y'}) \cr
G_0(y_1,y_1,{\bf \tilde y}|{\bf x'}) & 
L(y_1,{\bf \tilde y}|y_1',{\bf \tilde y'}) & G_0(y_1,y_1,{\bf \tilde
y}|u_1,{\bf y'}) & \cdots
& G_0(y_1,y_1,{\bf \tilde y}|u_M,{\bf y'}) \cr G_0(u_1,{\bf y}|{\bf x'}) &
G_0(u_1,{\bf y}|y_1', y_1',{\bf \tilde y'}) & K_1({\bf y}|{\bf y'}) & \cdots
& G_0(u_1,{\bf y}|u_M,{\bf y'}) \cr \vdots & \vdots & \vdots & \ddots & \vdots \cr
G_0(u_M,{\bf y}|{\bf x'}) & G_0(u_M,{\bf y}|y_1',y_1',{\bf \tilde y'}) &
G_0(u_M,{\bf y}|u_1,{\bf y'}) & \cdots & K_M({\bf y}|{\bf y'})} \right] \over
\det \left[ \matrix{ L(y_1,{\bf \tilde y}|y_1',{\bf \tilde y'}) &
G_0(y_1,y_1,{\bf \tilde y}|u_1,{\bf y'}) & \cdots
& G_0(y_1,y_1,{\bf \tilde y}|u_M,{\bf y'}) \cr G_0(u_1,{\bf
y}|y_1',y_1',{\bf \tilde y}) & K_1({\bf y}|{\bf y'}) & \cdots & G_0(u_1,{\bf
y}|u_M,{\bf y'}) \cr \vdots & \vdots & \ddots & \vdots \cr
G_0(u_M,{\bf y}|y_1',y_1',{\bf \tilde y'}) & G_0(u_M,{\bf y}|u_1,{\bf y'})
& \cdots &  K_M({\bf y}|{\bf y'})  } \right] }, 
\label{teljes-GF}
\end{equation}
where
\begin{eqnarray}
K_i({\bf y}|{\bf y'}) & = & -{1 \over N \kappa_i} \ \delta^{(N-1)} 
({\bf y}-{\bf y'}) +  G_0(u_i,{\bf y}|u_i',{\bf y'}),  
\label{Ki-alak}   \\
L(y_1,{\bf \tilde y}|y_1',{\bf \tilde y'}) & = & -{2 \over N(N-1) \lambda}
\delta(y_1-y_1') \  \delta^{(N-2)}({\bf \tilde y}-{\bf \tilde y'}) +
G_0(y_1,y_1,{\bf \tilde y}|y_1',y_1',{\bf \tilde y'}). 
\label{L-alak}
\end{eqnarray}
The variables ${\bf y} = (y_1,y_3,y_4,...,y_N)$  
and ${\bf \tilde y} = (y_3,y_4,...,y_N)$  (and the corresponding
primed ones)  appearing in the operators $L$ and $K_i$
are  $(N-1)$-component and $(N-2)$-component vectors, respectively.

At first sight the final form of the Green's function seems to be  
very complicated but the matrices have a simple structure.
The GF is again expressed as the ratio of two determinants.
The variables of the full GF appear only in the first row and first 
column of the matrix in the numerator. 
Omitting the first row and first column of the matrix in the numerator 
the remaining matrix is the same as that in the denominator.     
The matrix in the denominator has an obvious structure, too: 
the 1,1 element of the matrix,  $L$ 
contains only the strength of the interaction between the 
two particles, while omitting the first row and the first column
of this matrix it contains only the diagonal elements of the 
remaining matrix, i.e.\ the $K_i$ contain the strength of 
the impurities. The other matrix elements depend only on the Green's 
function $G_0$ of two noninteracting particles.   
 
In summary of this section, the energy levels of the interacting boson
systems with impurities are the poles of the Green function of the
systems which coincide with the zeros of the denominator 
in Eq.\ (\ref{teljes-GF}). This kind of form of the GF is a consequence of
the contact potential assumed for the interactions between particles
and the potential of the impurities. The denominator contains the 
operator $K$ and $L$ related to the impurities and the interactions of the 
bosons, respectively. In the special case when only the noninteracting 
$N>2$ bosons with impurities are considered, the denominator reduces to 
$\det K({\bf y}|{\bf y'})$ where $K$ is given by Eq.\ (\ref{Kmatrix}). 
The explicit form of these operators always available by using the 
{\em symmetrized} eigenfunctions of the noninteracting system. Then, these 
operators can be represented by an infinite dimension matrices. Note that,  
any basis set can be used (which is not necessaraly a symmetrized one) 
for evaluating the matrix elements of $K$ and $L$.  However, 
for numerical calculations of the matrices $K$ and $L$ one has to use 
a finite basis set. We would like to stress that in our method the
truncation of the infinite basis set is the {\em only} approximation
that is made.  
The advantage of our method against the direct diagonalization of the full
Hamiltonian is that in our approach the bosonic/fermionic feature of the
many-body systems is already  taken into account through the Green
function of the Hamiltonian $H_0$ of the noninteracting, impurity free
system. Then, it is ensured that the Green function of system of the 
interacting particles with impurities is also symmetrized 
according the nature of the particles. 
As we shall see below, the truncation introduce less error 
in the energy levels compared to that obtained by direct diagonalization
than the lack of this symmetrization.

\section{Spin-full fermions with interaction and impurities} 
\label{spin-fermion}

In this section  our spectral determinant method is generalized 
to an interacting many-body systems in which the particles can be
bosons or fermions with spins. 
The Green function can be derived in a way similar to the spinless 
bosons discussed in the previous sections. 
Note that for spinless fermions with the Dirac-delta interaction potential
the energy levels are the same as in the case of noninteracting
fermions.  

The GF now depends on the spin indices and for noninteracting particles 
GF is
\begin{equation}
G_0({\bf x}|{\bf x'},E)_{{\bf s}|{\bf s'}}=
\sum_{n}{\psi_n({\bf x})_{\bf s} \psi^*_n({\bf x'})_{\bf s'}\over E-E_n},
\end{equation}
where ${\bf s}=(s_1,s_2,...s_N)$ 
is a compact notation of the spins of the $N$ particles. 
The interacting part of the Hamiltonian again consist of two terms:
\begin{equation}
H_1({\bf x})_{{\bf s}|{\bf s'}}=\sum_{i=1}^M \kappa_i \sum_{p=1}^N 
\delta(x_p-u_i)\delta_{{\bf s},{\bf s'}}+\lambda
\sum_{{p,q=1}\atop{p<q}}^N 
\delta(x_p-x_q) \delta_{{\bf s},{\bf s'}},
\end{equation}
where all the terms are diagonal in the spin indices. Here $u_i$ and
$\kappa_i$ are  positions and the strength of the
impurity $i=1,\cdots, M$, and $\lambda$ is the strength of 
the interactions between  particles. 

Following the method developed in the previous sections we find that
the GF for the system of interacting particles (fermion or bosons) 
with spins and impurities is given by  
\begin{equation}
G({\bf x}|{\bf x'})_{{\bf s}|{\bf s'}} = {\det 
\left[ \matrix{G_0({\bf x}|{\bf x'})_{{\bf s}|{\bf s'}} & 
G_0({\bf x}|y_1',y_1',{\bf \tilde y'})_{{\bf s}|{\bf \bar s'}} & 
G_0({\bf x}|u_j,{\bf y'})_{{\bf s}|{\bf \bar s'}} \cr
G_0(y_1,y_1,{\bf \tilde y}|{\bf x'})_{{\bf \bar s}|{\bf s'}} & 
L(y_1,{\bf \tilde y}|y_1',{\bf \tilde y'})_{{\bf \bar s}|{\bf \bar s'}} & 
G_0(y_1,y_1,{\bf \tilde y}|u_j,{\bf y'})_{{\bf \bar s}|{\bf \bar s'}} \cr
G_0(u_i,{\bf y}|{\bf x'})_{{\bf \bar s}|{\bf s'}} & 
G_0(u_i,{\bf y}|y_1', y_1',{\bf \tilde y'})_{{\bf \bar s}|{\bf \bar s'}} & 
K_{i,j}({\bf y}|{\bf y'})_{{\bf \bar s}|{\bf \bar s'}}  } \right] \over 
\det \left[ \matrix{ L(y_1,{\bf \tilde y}|y_1',{\bf \tilde y'})_{{\bf \bar s}|{\bf \bar s'}} & 
G_0(y_1,y_1,{\bf \tilde y}|u_j,{\bf y'})_{{\bf \bar s}|{\bf \bar s'}} \cr 
G_0(u_i,{\bf y}|y_1',y_1',{\bf \tilde y})_{{\bf \bar s}|{\bf \bar s'}} & 
K_{i,j}({\bf y}|{\bf y'})_{{\bf \bar s}|{\bf \bar s'}}} \right] },
\label{fermion-GF}
\end{equation}
where
\begin{equation}
K_{i,j}({\bf y}|{\bf y'})_{{\bf \bar s}|{\bf \bar s'}} = -{1 \over N \kappa_i} \delta_{i,j} \delta^{(N-1)} ({\bf y}-{\bf y'}) \  \delta^{(N)}_{{\bf \bar s},{\bf \bar s'}} +  G_0(u_i,{\bf y}|u_j,{\bf y'})_{{\bf \bar s}|{\bf \bar s'}},
\end{equation}
and
\begin{equation}
L(y_1,{\bf \tilde y}|y_1',{\bf \tilde y'})_{{\bf \bar s}|{\bf \bar s'}} = -{2 \over N(N-1) \lambda} \delta(y_1-y_1') \ \delta^{(N-2)}({\bf \tilde y}-{\bf \tilde y'}) \ \delta^{(N)}_{{\bf \bar s},{\bf \bar s'}} + G_0(y_1,y_1,{\bf \tilde y}|y_1',y_1',{\bf \tilde y'})_{{\bf \bar s}|{\bf \bar s'}}.
\end{equation}
Here the same notations are  used for ${\bf y} = (y_1,y_3,y_4,...,y_N)$ and 
${\bf \tilde y} = (y_3,y_4,...,y_N)$
as in Eq.\ (\ref{teljes-GF}). The ${\bf s}, {\bf \bar s}$ denote 
the $N$ spin indices.

\section{The test of the spectral determinant method for bosons} 
\label{test}
In this section we apply our formalism in one dimension
using periodic boundary conditions to two, three and four interacting
spinless bosons without including impurities, as well as two bosons
with one impurity. 
The energy eigenvalues of these many-body systems can be found from 
the zeros of a SD derived from  GF formalism. 
In these systems one can calculate the exact energy levels from the
well known Bethe ansatz\cite{Bethe} solutions.  
Thus, our results can be compared with  the exact ones.
Using one-particle wave functions one
can build up the symmetrized many-body wave functions, and in this 
truncated basis the Hamiltonian of the system can be 
diagonalized numerically. We use the same number of one-particle
wave functions  in the SD method as in the diagonalization of the
Hamiltonian. If the number of one-particle wave functions used 
in the SD method is $p$, then for $N>2$ interacting particles 
one needs to find the zeros of a SD of size  
$p^{N-1}$.
On the other hand, in the diagonalization of the Hamiltonian using the 
same $p$ one-particle wave functions the size of the matrix that
should be diagonalized is 
$\left( \begin{array}{c}
p+N-1 \\
N
\end{array}
\right)
$.

It is interesting to mention that for two interacting particles the 
energy levels are {\it exactly\/} the same as those obtained from 
the Bethe ansatz solutions. The details of the derivation is given in 
 Appendix \ref{2boson}. 
In sections \ref{3boson} and \ref{4boson} the derivation of 
the equations for the energy eigenvalues and the numerical results are 
presented for three and four interacting particles. 
The comparison of the SD method with direct diagonalization shows 
that the SD method leads to an order of magnitude improvement in the accuracy 
even in the worst case.

\subsection{Three interacting bosons  in one dimension}
\label{3boson}

Now the method developed in section \ref{int+impuri} is
applied for $N=3$ interacting particles in a one-dimensional box 
with periodic boundary conditions. 
To determine the Green's function $G_0$  for the case of three particles, 
it is again convenient to use the one-particle wave functions and the
corresponding eigenvalues given by Eqs.\ (\ref{1wave}) and (\ref{E_n}). 
Hereafter we shall use the length unit $a=1$ for the size of the box. 
Then the Green's function $G_0$ is given by
\begin{equation}
G_0(x_1,x_2,x_3|x_1',x_2',x_3') = \sum \limits_{\{n_1n_2n_3\}} 
{\psi_{n_1n_2n_3}(x_1,x_2,x_3) \ \psi^*_{n_1n_2n_3}(x_1',x_2',x_3')
\over E-E_{n_1}-E_{n_2}-E_{n_3}}, 
\end{equation} 
where the $\psi_{n_1n_2n_3}(x_1,x_2,x_3)$ are the eigenfunctions of 
the system of three noninteracting particles, and 
$n_1,n_2,n_3=0,\pm 1,\pm 2, \cdots$. 
For bosons these wavefunctions 
can be built up from the one-particle wave functions given by 
Eq.\ (\ref{1wave}). 
Finally, the GF can be simplified as 
\begin{equation}
G_0(x_1,x_2,x_3|x_1',x_2',x_3') = {1 \over 3!} \sum \limits_{n_1n_2n_3} 
\sum \limits_{P \in S_3} {\psi_{n_1}(x_1) \ \psi_{n_2}(x_2) \ \psi_{n_3}(x_3) 
\ \ \psi^*_{n_{P(1)}}(x_1') \ \psi^*_{n_{P(2)}}(x_2') \ \psi^*_{n_{P(3)}}
(x_3') \over E-E_{n_1}-E_{n_2}-E_{n_3}},
\label{G0-3} 
\end{equation}
where $S_3$ denotes the permutation group of the numbers $1,2,3$, and 
$P$ is an element of the group, while $P(i)$ is the number 
in the $i$th place in  permutation $P$.  

The energy levels of the system are the poles of the GF given 
by Eq.\ (\ref{teljes-GF}). From the form of the GF we can see that obtaining 
the poles is equivalent to finding the solutions of the equation 
$\det L =0$ for $E$, where $L$ is given by Eq.\ (\ref{L-alak}), i.e.:
\begin{equation}
\det L(y,x|y',x') = \det \left( -{1 \over 3\lambda} \ \delta(y-y') \ 
\delta(x-x') + G_0(y,y,x|y',y',x';E) \right) = 0,
\label{L3=0}
\end{equation}
where $\lambda$ is the strength of the
intercation between bosons.
Using Eqs.\ (\ref{1wave}) and  (\ref{G0-3}) the second term in $\det L$ 
becomes
\begin{equation}
G_0(y,y,x|y',y',x') = {1 \over 3} \sum 
\limits_{n_1n_2n_3} 
\frac{e^{2 \pi i \left[(n_1+n_2)(y-y')+n_3(x-x')\right]} 
+ 2 e^{2 \pi i \left[ n_1(y-x')+ n_2(x-y')+n_3(y-y')\right]}}
{ E-E_{n_1}-E_{n_2}-E_{n_3}}.
\end{equation}

The determinant of the operator $L$ in Eq.\ (\ref{L3=0}) can be rewritten 
as a determinant of an infinite sized matrix using any convenient basis 
(which  is not necessarily that of symmetrized wave functions). In our
calculation we chose
\begin{equation} 
\varphi_{kl} (y,x) = e^{2 \pi i (ky+lx)}.
\label{basis}
\end{equation}
Thus, the operator $L$ in this basis can be given by
\begin{equation}
L(kl|k'l') = \int \varphi^*_{kl} (y,x) \ L(y,x|y',x') \ \varphi_{k'l'} 
(y',x') \ dy \ dx \ dy'\ dx'.
\end{equation}
After some tedious but straightforward algebra one can find that
\begin{equation}
L(kl|k'l') = -{1\over 3\lambda}\, \delta_{k,k'}\delta_{l,l'}
+\frac{1}{3}\, \left(\frac{2\, \delta_{k+l,k'+l'}}{ E-E_{l'}-E_l-E_{k-l'}}
+\, \sum_n \frac{\delta_{k,k'}\delta_{l,l'}}{E-E_{k-n}-E_n-E_l}
\right).
\end{equation}
The sum in this equation has a similar structure to 
Eq.\ (\ref{eq-2p}), therefore using the identities given 
in Eq.\ (\ref{azonossag}) the summation can be carried out yielding 
\begin{equation}  
L(k,l|k',l')=\left\{
\begin{array}{ll}
\delta_{k,k'} \delta_{l,l'} 
\left(-{1 \over 3\lambda}+{1 \over 12 \pi z} 
\cot {\pi \over 2} z \right) + A(E;k,l|k',l'), 
\,\,\,\, {\rm for}\,\,  k \,\, {\rm even},\\[1ex]
-\delta_{k,k'} \delta_{l,l'} 
\left({1 \over 3\lambda}+{1 \over 12 \pi z}
\tan {\pi \over 2} z \right)  + A(E;k,l|k',l'),
\,\,\,\, {\rm for} \,\, k \,\, {\rm odd},
\end{array} 
\right. 
\end{equation}
where
\begin{eqnarray}
A(E;k,l|k',l') & = & {2 \over 3}\,
{\delta_{k+l,k'+l'}
\over E-4 \pi^2\left[l^2+l'^2+(k-l')^2\right]}, \nonumber \\
z & = & \sqrt{{E \over 2 \pi^2}-2 l^2 -k^2 }.
\end{eqnarray}

One can see that choosing an appropriate basis it is possible 
to express all the matrix elements of the operator $L$ in a closed form. 
We now solve Eq.\ (\ref{L3=0}) numerically to find the energy levels 
from our method and compare them with the Bethe ansatz solutions determined  
in Eqs.\ (\ref{Bethe-eq}) and (\ref{Bethe-E}).

In numerical calculations we have to truncate the matrix $L(k,l|k',l')$. 
To form the basis  $\varphi_{kl}(x,y)$ given in (\ref{basis}) 
we chose $2m+1$ one-particle states 
$\psi_k(x)$ ($k=0,\pm 1,\pm 2,\cdots,\pm m$) 
as given in Eq.\ (\ref{1wave}). 
Therefore, we have ${(2m+1)}^2$ different basis functions 
$\varphi_{kl}(x,y)$ with $k,l=0,\pm 1,\pm 2,\cdots,\pm m$. 
Thus, the infinite sized matrix 
$L(k,l|k',l')$ is truncated to a ${(2m+1)}^2 \times {(2m+1)}^2$ matrix.   

\begin{figure}[hbt]
{\centerline{\leavevmode \epsfxsize=8.5cm \epsffile{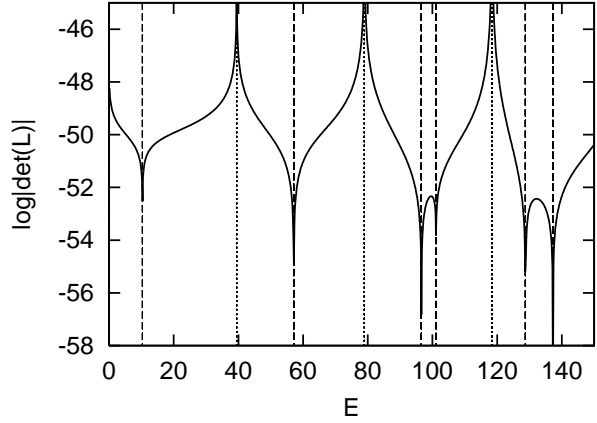 }}}
\caption{The logarithm of the absolute value of $\det L(k,l|k',l')$ 
as a function of $E$ for $N=3$ interacting particles (solid line). 
The strength of the interactions is $\lambda=4.0$. 
The number of one-particle wave functions was $2m+1=7$. 
Including the interactions, the energy levels, i.e.\ the position of
the zeros obtained from the Bethe ansatz solution are shown by dashed
vertical lines.   
For clarity, we indicate the energy
levels of noninteracting particles  with dotted vertical lines 
($E_k=4\pi^2k^2$, where $k=0,1,2,\cdots$). 
\label{3b-abra}}
\end{figure}

The logarithm of the absolute value of $\det L(k,l|k',l')$ is 
plotted as a function of the energy $E$ in Fig.\ \ref{3b-abra}. 
In this plot $2m+1=7$ one-particle wave functions were used. 
The zeros of $\det L(k,l|k',l')$ (i.e., in the plot at the energy values
where the logarithm of $\det L(k,l|k',l')$ goes to $-\infty$) 
give the energy levels of the system. One can see from the figure that 
these energy values agree very well with those found from the Bethe
ansatz solution. Note that the $\det L(k,l|k',l')$ has poles at the
energy levels of the noninteracting system. This can be seen clearly
from the figure.   
 
\begin{figure}[hbt]
{\centerline{\leavevmode \epsfxsize=8.5cm \epsffile{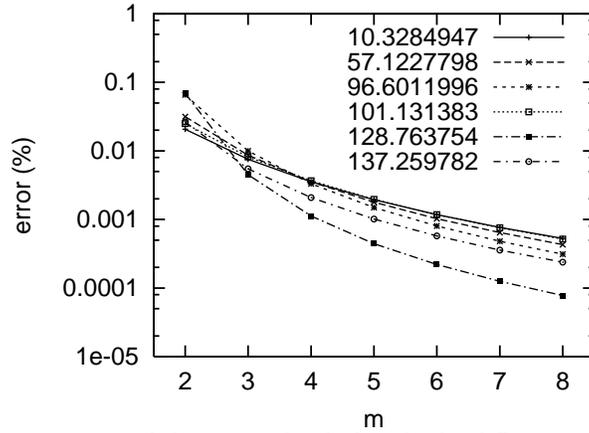 }}}
\caption{The relative errors (in percentage) of the energy levels
found (the different energy levels are indicated by different symbols) 
from the SD method using $2m+1$ one-particle 
wave functions as a function of $m$ for the first six energy
levels. The strength of the interactions 
is $\lambda=4.0$.
\label{error-abra}}
\end{figure}

In Fig.\ \ref{error-abra} the relative errors of the first six 
energy levels obtained from Bethe ansatz solutions and our SD method 
using $2m+1$ one-particle wave functions are plotted as a function of $m$.
One can see that the relative errors decrease with increasing $m$. 
It implies that including only a few number of one-particle
wave functions in our SD method is enough to get a very satisfactory 
agreements with the exact energy levels obtained by Bethe ansatz solutions. 

\begin{figure}[h]
{\centerline{\leavevmode \epsfxsize=8.5cm \epsffile{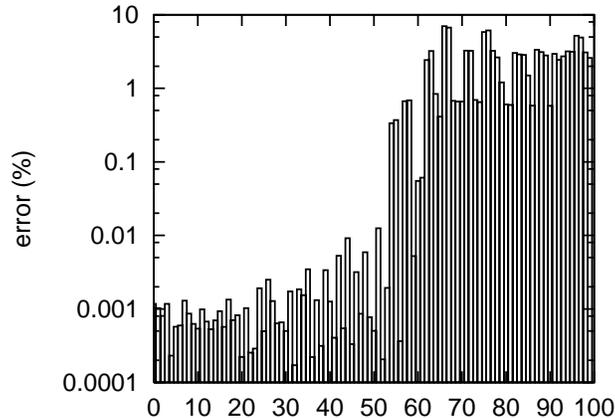 }}}
\caption{The relative errors (in percentage) of the the first $100$ 
energy levels found from the SD method using $2m+1=19$ one-particle 
wave functions. The strength of the interactions is $\lambda=4.0$.
\label{3f-abra}}
\end{figure}

In Fig.\ \ref{3f-abra} we plotted the relative errors of the energy 
levels found from the SD method for the first $100$ energy
levels. 

Diagonalization of the Hamiltonian can be performed by using the same 
one-particle wave functions as in the SD method. 
In Table \ref{tabla-b-3-GF} we have listed the first few energy levels
obtained from Bethe ansatz solutions, diagonalization of the
Hamiltonian of the system and the SD method. 
It can be seen from Table \ref{tabla-b-3-GF} that the relative errors 
of the energy levels 
obtained from the SD method are an order of magnitude smaller than 
those of the diagonalization of the Hamiltonian. 

One can see that the errors of the first 50 energy levels are
less than $0.01 \%$, in spite of the fact that only $9$ one-particle 
wave functions were used. 
It is surprising that only a relatively small
number of one-particle wave functions needs to be used for obtaining a 
rather accurate result. 
The accuracy breaks down after the first $50$ energy levels. This is
the point where the unperturbed energy levels corresponding to the
calculated levels start missing in case of 9 one-particle base
functions.
\begin{table}[hbt]
\begin{tabular}{||c|c|c|c|c|c|c||} \hline
\multicolumn{1}{||c} {Bethe ansatz} & 
\multicolumn{3}{|c} {Diagonalization}  &
\multicolumn{3}{|c||} {Spectral determinant method} \\ \hline 
$E$          & $m=2$    &  $m=3$  & $m=4$  & $m=2$  & $m=3$ & $m=4$ \\ \hline 
10.3284947 &   10.7066 (3.7) &   10.5994 (2.6) &   10.5403 (2.05) &   10.3264 (0.02) &  10.3277 (0.008) &   10.3281 (0.004)  \\ \hline  
57.1227798 &   57.9505 (1.4) &   57.6827 (1.0) &   57.5488 (0.75) &   57.1508 (0.05) &  57.1196 (0.006) &   57.1508 (0.049)  \\ \hline  
96.6011996 &   97.4767 (0.9) &   97.1866 (0.6) &   97.0426 (0.46) &   96.5507 (0.05) &  96.6507 (0.051) &   96.6507 (0.051)  \\ \hline  
101.1313830 &  102.8964 (1.7) &  102.7003 (1.6) &  102.5885 (1.44) &  101.1061 (0.03) & 101.1231 (0.008) &  101.1277 (0.004)  \\ \hline  
128.7637540 &  129.3888 (0.5) &  129.1411 (0.3) &  129.0332 (0.21) &  128.7006 (0.05) & 128.7584 (0.004) &  128.7756 (0.009)  \\ \hline  
137.2597820 &  137.7723 (0.4) &  137.4139 (0.1) &  137.2811 (0.02) &  137.2505 (0.01) & 137.3005 (0.030) &  137.2630 (0.002)  \\ \hline  
176.7382020 &  177.5799 (0.5) &  177.0631 (0.2) &  176.8474 (0.06) &  176.6504 (0.05) & 176.7504 (0.007) &  176.7340 (0.002)  \\ \hline  
219.5666420 &  222.0227 (1.1) &  221.2907 (0.8) &  221.1095 (0.70) &  212.7003 (3.13) & 219.5003 (0.030) &  219.6003 (0.015)  \\ \hline  
220.2550290 &  222.4741 (1.0) &  221.8307 (0.7) &  221.6792 (0.65) &  220.1503 (0.05) & 220.2378 (0.008) &  220.3003 (0.021)  \\ \hline  
254.5148790 &  256.0261 (0.6) &  255.0061 (0.2) &  254.8096 (0.12) &  252.7502 (0.69) & 254.4002 (0.045) &  254.5502 (0.014)  \\ \hline  
\end{tabular}
\caption{The energy levels from Bethe ansatz, diagonalization of the
Hamiltonian and the SD method. In the diagonalization method we used the
same one-particle wave functions as in the SD method. The percentage 
errors are indicated in brackets. The strength of 
the interactions is $\lambda=4.0$.
\label{tabla-b-3-GF}}
\end{table}

\subsection{Four interacting bosons in one dimension} \label{4boson}
 
In this subsection the equation for the energy levels is derived 
for $N=4$ interacting particles in a one-dimensional 
box with periodic boundary conditions. 
The necessary steps to find this equation are similar to
those in the previous subsection. In fact, the SD method can be
generalized straightforwardly to any number of interacting particles, 
although the algebra becomes more complicated for $N > 4$.    
The energy levels for  $N=4$ particles 
are the zeros of the determinant of operator $L$ given 
in Eq.\ (\ref{L-alak}).

The first few energy levels obtained from solving are listed 
in Table \ref{tabla-4-GF}. The strength of the
interactions  is $\lambda=4.0.$ 
For the sake of comparison the exact energy levels obtained from the
Bethe ansatz solution and the results found from the direct
diagonalization of the Hamiltonian are also listed in the table. 
Both in the SD method  and in the diagonalization method the same $2m+1$ 
one-particle wave functions were used. 
The percentage errors with respect to the exact Bethe ansatz solutions 
are indicated in brackets. 

From the table one can see that the relative errors using the SD method 
are more than an order of magnitude smaller than those obtained from direct
diagonalization. In particular, using a rather small number of
one-particle wave functions, $2m+1=9$, the ground state energy 
is much more accurate ($0.01\%$) compared to those obtained from the 
diagonalization method.

Note that to find the energy levels with the above indicated errors
only the determinant of a $9\times 9$ matrix had to be calculated. 
This suggests that in numerical calculations the SD method is much more 
efficient than the conventional diagonalization method.
Comparing the two methods one realizes that 
although in our method more algebra is needed than in the diagonalization 
methods, less numerical effort is required to obtain the values of 
the energy levels accurately. 
It is important to note that the expression for determining
the energy eigenvalues becomes more complicated with increasing the number of 
interacting particles but the necessary steps are quite straightforward
and a quite simple algorithm can be given. 

\begin{table}[hbt]
\begin{tabular}{|c|c|c|c|c|c|c|} \hline
\multicolumn{1}{||c} {Bethe ansatz} & 
\multicolumn{3}{|c} {Diagonalization}  &
\multicolumn{3}{|c||} {Spectral determinant method} \\ \hline 
$E$          & $m=2$    &  $m=3$  & $m=4$  & $m=2$  & $m=3$ & $m=4$ \\ \hline 
  20.8016 &      0.00 (100.0) &     21.34 (2.6) &     21.22 (2.01) &   20.7904 (0.05) &  20.7976 (0.02) &   20.7997 (0.009)  \\ \hline  
  70.8207 &     72.18 (1.9) &     71.74 (1.3) &     71.51 (0.97) &   70.7561 (0.09) &  70.8031 (0.02) &   70.8135 (0.010)  \\ \hline  
 113.7688 &    115.36 (1.4) &    114.83 (0.9) &    114.56 (0.69) &  113.6250 (0.13) & 113.7380 (0.03) &  113.7576 (0.010)  \\ \hline  
 118.4289 &    120.12 (1.4) &    119.59 (1.0) &    119.31 (0.74) &  118.3495 (0.07) & 118.4014 (0.02) &  118.4168 (0.010)  \\ \hline  
 149.7776 &    151.31 (1.0) &    150.78 (0.7) &    150.51 (0.49) &  149.4374 (0.23) & 149.7434 (0.02) &  149.7660 (0.008)  \\ \hline  
 158.5862 &    160.32 (1.1) &    159.78 (0.8) &    159.49 (0.57) &  158.3920 (0.12) & 158.5536 (0.02) &  158.5722 (0.009)  \\ \hline  
 178.7153 &    179.98 (0.7) &    179.47 (0.4) &    179.25 (0.30) &  178.3744 (0.19) & 178.6877 (0.02) &  178.7084 (0.004)  \\ \hline  
 191.2125 &    193.25 (1.1) &    192.43 (0.6) &    192.08 (0.46) &  191.0778 (0.07) & 191.1803 (0.02) &  191.2018 (0.006)  \\ \hline  
 195.3592 &    197.32 (1.0) &    196.46 (0.6) &    196.20 (0.43) &  195.2056 (0.08) & 195.3277 (0.02) &  195.3468 (0.006)  \\ \hline  
 237.5430 &    240.82 (1.4) &    239.00 (0.6) &    238.59 (0.44) &  237.7028 (0.07) & 237.3591 (0.08) &  237.5022 (0.017)  \\ \hline  
 238.2730 &    241.38 (1.3) &    239.68 (0.6) &    239.31 (0.43) & - & 238.2151 (0.02) &  238.2512 (0.009)  \\ \hline  
 276.3426 &    279.69 (1.2) &    277.84 (0.5) &    277.40 (0.38) &  274.9712 (0.50) & 275.8985 (0.16) &  276.2966 (0.017)  \\ \hline  
 277.9374 &    280.87 (1.1) &    279.30 (0.5) &    278.97 (0.37) &  277.7805 (0.06) & 277.8731 (0.02) &  277.9146 (0.008)  \\ \hline  
 281.8038 &    281.29 (0.2) &    283.42 (0.6) &    282.94 (0.40) &  285.7567 (1.40) & 281.5560 (0.09) &  281.7657 (0.014)  \\ \hline  
\end{tabular}
\caption{The exact energy levels obtained from the Bethe ansatz 
and our numerical results using  $2m+1$ one-particle wave functions.
In the diagonalization method we used the
same one-particle wave functions as in the SD method.  
The relative errors in percentage are indicated in brackets.
The strength of the interactions is $\lambda=4.0.$
\label{tabla-4-GF}}
\end{table}

\section{Two interacting bosons with one impurity 
in one dimension} \label{2p+impuri} 

As a further illustration of the SD method, 
in this subsection we consider a
system including two particles and an impurity in a one-dimensional box
with periodic boundary conditions. 
In this case no exact results, such as Bethe ansatz solutions, 
are known for the energy levels. The Hamiltonian of the system is 
$H=H_0 + H_1$, where $H_0$ is given by Eq.\  (\ref{H0-op}).
The interaction Hamiltonian of the system can be written as
\begin{equation}
H_1(x_1,x_2)= \lambda \delta(x_1-x_2)
+\kappa \left[\delta \left(x_1-u \right)+
\delta \left(x_2-u\right) \right],
\end{equation}
where the strength of the interaction between the two particles and 
the strength of the potential for the impurity are denoted by
$\lambda$ and $\kappa$, respectively. The impurity is located at $u$.  
The GF of the system is given by Eq.\ (\ref{teljes-GF}) with $M=1$. 
The energy levels of the system are the roots of the
denominator in (\ref{teljes-GF}), which in our case, can be written as  
\begin{equation}
\det \left[\matrix{L(y,y';E) & G_0(y,y|y',u;E) \cr G_0(y,u|y',y';E)
& K(y,y';E) } \right] = 0, 
\label{det-int-imp} 
\end{equation}
where
\begin{eqnarray}
L(y,y') & = & -{1 \over \lambda} \delta(y-y')+G_0(y,y|y',y'), \\
K(y,y') & = & -{1 \over 2 \kappa} \delta(y-y')+G_0(y,u|y',u).
\end{eqnarray}
The Green's function $G_0(x_1,x_2|x_1',x_2')$ of the 
Hamiltonian $H_0$ can be written as
\begin{eqnarray}
\lefteqn{G_0(x_1,x_2|x_1',x_2') = {1 \over 2!} \sum \limits_{n_1n_2} 
\sum \limits_{P \in S_2} {\psi_{n_1}(x_1) \ \psi_{n_2}(x_2) \ \ 
\psi^*_{n_{P(1)}}(x_1') \ \psi^*_{n_{P(2)}}(x_2') \over E-E_{n_1}-E_{n_2}} 
}\nonumber \\
&& ={1 \over 2} \left( \sum \limits_{n,m} {\psi_n(x_1) \ \psi_m(x_2) \ \ 
\psi^*_n(x_1') \ \psi^*_m(x_2') \over E-E_n-E_m} +  \sum \limits_{n,m} 
{\psi_n(x_1) \ \psi_m(x_2) \ \ \psi^*_m(x_1') \ \psi^*_n(x_2') \over 
E-E_n-E_m} \right),
\end{eqnarray}
where $\psi_n(x)$ and $E_n$ are given by Eqs.\ (\ref{1wave}) and (\ref{E_n}), 
respectively (with $a=1$).
In Eq.\ (\ref{det-int-imp}) the four matrix elements 
contain the two-particle Green's function 
$G_0(x_1,x_2|x_1',x_2')$ with different arguments. 
The four matrix elements are operators but for numerical calculations 
we need to represent them by
matrices. This can be done by using a set of orthogonal wave functions:
\begin{equation}
\varphi_{k} (y) = e^{2 \pi i k (y-u)},
\end{equation}
where the extra phase $e^{-2 \pi i u} $ is introduced for the sake 
of convenience.
Then, in this basis, the Green's function $G_0$ appearing in 
the matrix elements $1,1$, $1,2$, $2,1$ and $2,2$ of 
Eq.\ (\ref{det-int-imp})  becomes 
\begin{eqnarray}
G_0^{(11)}(k|l) & = &
\delta_{k,l} \sum_n  {1\over E-E_n-E_{k-n}}, \nonumber \\
G_0^{(12)}(k|l) & = & {1\over E-E_l-E_{k-l}}, \nonumber \\
G_0^{(21)}(k|l) & = & {1\over E-E_k-E_{l-k}} = G_0^{(1,2)}(l|k),  \\
G_0^{(22)}(k,l) & = &
{1 \over 2} \, \left({1\over E-E_k-E_l} + \delta_{k,l} 
\sum_n {1\over E-E_k-E_n} \right). \nonumber 
\end{eqnarray} 
The sum over $n$ in $G_0^{(11)}(k|l)$ can be carried out by 
using the identities  Eq.\ (\ref{azonossag}), 
while the summation in $G_0^{(22)}(k|l)$ can be performed  using the 
identity\cite{Abramowitz} 
\begin{equation}
\sum_{n=-\infty}^{\infty} \frac{1}{z^2 - n^2} = \frac{\pi \cot \pi z }{z}.
\end{equation}
Finally, the matrix representation of Eq.\ (\ref{det-int-imp}) becomes
\begin{equation}
\det \left[ \begin{array}{cc}
{M}^{\left(11\right)} & {M}^{\left(12\right)} \\[2ex]
{M}^{\left(21\right)} & {M}^{\left(22\right)} 
\end{array}
\label{imp-det}
\right] =0,
\end{equation}
where the four elements of the hypermatrix  are 
\begin{eqnarray} 
{M}^{\left(11\right)}_{k,l} & 
= & \delta_{k,l}\left( -{1 \over \lambda}+ S_k \right), 
\nonumber \\[2ex]
{M}^{\left(12\right)}_{k,l} & = & {1\over E-4\pi^2\left[l^2+(k-l)^2\right]}, 
\,\,\,\,\,\, {M}^{\left(21\right)}_{k,l} = {M}^{\left(12\right)}_{l,k},  
\\[2ex]
{M}^{\left(22\right)}_{k,l} & = & \delta_{k,l} \left( -{1 \over 2 \kappa} 
+ {1 \over 4 \pi z_2} \cot \pi z_2 
\right)+ {1\over E-4\pi^2(k^2+l^2)}, \nonumber 
\end{eqnarray}
with
\begin{eqnarray}
S_k & = & \left\{ \begin{array}{ll}
\frac{1}{4 \pi z_1}\cot \left( \frac{\pi}{2}z_1\right)
\,\,\,\, \mbox{for}\,\,\,\, k \,\,\, \mbox{even,}  \\[1ex]
-\frac{1}{4 \pi z_1}\tan \left( \frac{\pi}{2}z_1 \right)
\,\,\,\, \mbox{for}\,\,\,\, k \,\,\, \mbox{odd,} 
\end{array} \right.   \\
z_1 & = & \sqrt{{E\over 2 \pi^2}-k^2} \,\,\,\,\, \mbox{and} \,\,\,
z_2  =  \sqrt{{E \over 4 \pi^2} -k^2}.
\end{eqnarray}
Note that to calculate numerically the determinant in 
Eq.\ (\ref{imp-det}) it is useful to apply an equivalent form of 
the general matrix identity 
for hypermatrices given in Eq.\ (\ref{hyperm}) 
\begin{equation}
{\rm det} \left(
\begin{array}{cc} 
a & b \\
c & d 
\end{array}
\right ) = {\rm det} (a)\  {\rm det}
\left(d - c a^{-1}b \right )
\label{hyper}
\end{equation} 
(which holds for arbitrary square matrices $a$ and $d$, 
with $\det a \neq 0$). Then, the dimensions of the 
minors are half of the original one. 
Since the matrix ${L}^{(11)}_{k,l}$ is diagonal, it is easy to calculate 
its inverse.  

The first few energy levels obtained from solving  Eq.\ (\ref{imp-det}) 
numerically are listed in Table \ref{tabla-int-imp-GF}. 
The strength of the interactions and the strength of potential 
for the impurity are $\lambda=3.0$ and $\kappa = 2.0$, respectively.

\begin{table}
\begin{tabular}{|c|c|c|c|c|c|c|} \hline
\multicolumn{1}{||c} {`Exact' $E$ (4 digits)} & 
\multicolumn{3}{|c} {Diagonalization}  &
\multicolumn{3}{|c||} {Spectral determinant method} \\ \hline 
$E$          & $m=2$    &  $m=3$  & $m=4$  & $m=2$  & $m=3$ & $m=4$ \\ \hline
   6.1175 &    6.3153 (3.2) &    6.2583 (2.3) &    6.2262 (1.78) &    6.1164 (0.018) &   6.1171 (0.0067) &    6.1173 (0.0032)  \\ \hline  
  46.8629 &   47.1405 (0.6) &   47.0458 (0.4) &   47.0002 (0.29) &   46.8624 (0.001) &  46.8628 (0.0002) &   46.8629 (0.0000)  \\ \hline  
  50.4776 &   50.8832 (0.8) &   50.7582 (0.6) &   50.6908 (0.42) &   50.4695 (0.016) &  50.4750 (0.0051) &   50.4765 (0.0023)  \\ \hline  
  82.9012 &   83.0716 (0.2) &   82.9998 (0.1) &   82.9746 (0.09) &   82.9007 (0.001) &  82.9010 (0.0001) &   82.9011 (0.0001)  \\ \hline  
  85.5893 &   85.8157 (0.3) &   85.7852 (0.2) &   85.7338 (0.17) &   85.5869 (0.003) &  85.5891 (0.0003) &   85.5893 (0.0001)  \\ \hline  
  91.2389 &   91.7645 (0.6) &   91.5606 (0.4) &   91.4876 (0.27) &   91.2224 (0.018) &  91.2350 (0.0043) &   91.2373 (0.0018)  \\ \hline  
 165.5271 &  165.9033 (0.2) &  165.7284 (0.1) &  165.6938 (0.10) &  165.5104 (0.010) & 165.5263 (0.0005) &  165.5269 (0.0001)  \\ \hline  
 169.3990 &  170.0360 (0.4) &  169.7856 (0.2) &  169.6524 (0.15) &  169.3588 (0.024) & 169.3938 (0.0030) &  169.3972 (0.0011)  \\ \hline  
 203.1501 &  203.5057 (0.2) &  203.3393 (0.1) &  203.2789 (0.06) &  201.3532 (0.885) & 203.1501 (0.0000) &  203.1501 (0.0000)  \\ \hline  
 206.9552 &  207.5140 (0.3) &  207.2107 (0.1) &  207.1694 (0.10) &  207.1683 (0.103) & 206.9520 (0.0015) &  206.9547 (0.0002)  \\ \hline  
 207.1812 &  207.5666 (0.2) &  207.4673 (0.1) &  207.3835 (0.10) &  209.2883 (1.017) & 207.1799 (0.0006) &  207.1811 (0.0001)  \\ \hline  
 210.9868 &  211.9115 (0.4) &  211.4817 (0.2) &  211.3105 (0.15) & - & 210.9672 (0.0093) &  210.9816 (0.0024)  \\ \hline  
\end{tabular}
\caption{The `exact' energy levels (results to four significant figures 
using 61 one-particle wave functions) 
and our numerical results obtained from
Eq.\  (\ref{imp-det}) using  $2m+1$ one-particle wave functions.
In the diagonalization method we used the
same one-particle wave functions as in the SD method. 
The percentage errors are indicated in brackets.
The strength of the interactions and the strength of potential 
for the impurity are $\lambda=3.0$ and $\kappa = 2.0$, respectively.
\label{tabla-int-imp-GF}}
\end{table}

No Bethe ansatz type solution is known when the system includes 
impurity. The `exact' energy eigenvalues are calculated by  
increasing the number of one-particle wave functions until  
all the energy levels listed in the first column of the table 
are converged to four significant figures. 
To reach this convergence, $61$ one-particle wave functions 
were used. Then the errors of the energy eigenvalues found from the SD
method and from the direct diagonalization of the Hamiltonian are
compared using the same number of one-particle wave functions. 
As it is seen from Table \ref{tabla-int-imp-GF}, our method usually gives 
again an order of magnitude more accurate results than the 
diagonalization of the Hamiltonian.

\section{Three interacting fermions with one impurity 
in one dimension and calculation of the persistent current} 
\label{3pfermion+impuri} 

In this section our SD method is applied for a more complicated
problem. We consider the one dimensional ring including  
$N=3$ interacting 1/2 spin-fermions. The external magnetic field is
applied along the axis perpendicular to the plane of the ring. 
In addition, a single impurity is also included in an arbitrary position.
The persistent current is the derivative of the ground
state with respect to the flux of the magnetic field 
inside the ring\cite{Buttiker1,Cheung}:
\begin{equation}
I = -c\, \frac{d E(\Phi)}{d \Phi},
\end{equation}
where $E(\Phi)$ is the energy of the ground state of the system 
with flux $\Phi$ through the ring.
The ground state energy is the smallest poles of the GF for fermion 
systems given by Eq.\ (\ref{fermion-GF}) in section \ref{spin-fermion}. 

The GF is constructed from the one-particle wavefunctions:
\begin{equation}
\psi_{n,\sigma}(x)_s=e^{2\pi i (n+\Phi) x} \sigma_s,
\label{fermion-one-partic}
\end{equation}
where $\sigma = \left|\uparrow \right>, \left|\downarrow \right>$, 
$s=1,2$ denoting the spinor index of the spin state $\sigma$ 
and $\Phi$ is the magnetic flux in units of flux quantum $h/e$. 
The corresponding eigenvalues of the free one-electron Hamiltonian are 
$E_n=4\pi^2{\left(n+\Phi \right)}^2$ with $n=0,\pm 1,\pm 2,\cdots$.

The unperturbed GF appearing in Eq.\ (\ref{fermion-GF}) can be
constructed from the above given one-particle wave functions. For
fermions the symmetrized form is: 
\begin{equation}
G_0= {1\over 3!} \sum_{P\in S_3} 
\sum_{n_1 n_2 n_3 \atop {\sigma_1 \sigma_2 \sigma_3}}
(-1)^P { \psi_{n_1,\sigma_1}(x_1)_{s_1} \psi_{n_2,\sigma_2}(x_2)_{s_2} 
\psi_{n_3,\sigma_3}(x_3)_{s_3} \psi^*_{n_{P(1)},\sigma_{P(1)}}(x_1')_{s_1'} 
\psi^*_{n_{P(2)},\sigma_{P(2)}}(x_2')_{s_2'} \psi^*_{n_{P(3)},\sigma_{P(3)}}
(x_3')_{s_3'}
\over E-E_{n_1}-E_{n_2}-E_{n_3}},
\end{equation}
where $S_3$ denotes the permutation group as given after 
Eq.\ (\ref{G0-3}).
The denominator of the GF in (\ref{fermion-GF}) can be transformed into
a matrix form by using the non-symmetrized product of the one-particle 
wave functions as it was done in the case of bosons.
The determinant of this matrix takes the following 2 by 2 hypermatrix form:
\begin{equation}
\det \left[\matrix{M^{(11)} & M^{(12)} \cr M^{(21)} & M^{(22)}  } \right],
\label{polus-eq-fermion}
\end{equation}
where  the $p=(k,l,s_1,s_2,s_3),\, q=(k',l',s_1',s_2',s_3')$ matrix
element of the matrices $M$ is given by   
\begin{eqnarray}
M_{pq}^{(11)}&=&-{1\over 3 \lambda} \delta_{k,k'} \delta_{l,l'} 
\delta_{s_1,s_1'} \delta_{s_2,s_2'} \delta_{s_3,s_3'}+
{1\over 6} \sum_{P\in S_3} 
\sum_{n_1,n_2,n_3} {\delta_{n_1+n_2,k} \delta_{n_3,l} 
\delta_{n_{P(1)}+n_{P(2)},k'} \delta_{n_{P(3)},l'}
\over E-E_{n_1}-E_{n_2}-E_{n_3}} \delta_{s_{P(1)},s_1'} 
\delta_{s_{P(2)},s_2'} \delta_{s_{P(3)},s_3'}, \\[1ex]
M_{pq}^{(12)}&=& {1\over 6} \sum_{P\in S_3} \sum_{n_1,n_2,n_3} 
{\delta_{n_1+n_2,k} \delta_{n_3,l} \delta_{n_{P(2)},k'} \delta_{n_{P(3)},l'}
\over E-E_{n_1}-E_{n_2}-E_{n_3}} 
\delta_{s_{P(1)},s_1'} \delta_{s_{P(2)},s_2'} \delta_{s_{P(3)},s_3'},  \\[1ex]
M_{pq}^{(21)}&=& {1\over 6} \sum_{P\in S_3} \sum_{n_1,n_2,n_3} 
{\delta_{n_2,k} \delta_{n_3,l} \delta_{n_{P(1)}+n_{P(2)},k'} 
\delta_{n_{P(3)},l'}\over E-E_{n_1}-E_{n_2}-E_{n_3}} \delta_{s_{P(1)},s_1'} 
\delta_{s_{P(2)},s_2'} \delta_{s_{P(3)},s_3'}, \\[1ex]
M_{pq}^{(22)}&=&-{1\over 3 \kappa} \delta_{k,k'} \delta_{l,l'} 
\delta_{s_1,s_1'} \delta_{s_2,s_2'} \delta_{s_3,s_3'}+
{1\over 6} \sum_{P\in S_3}\sum_{n_1,n_2,n_3} {\delta_{n_2,k} \delta_{n_3,l} 
\delta_{n_{P(2)},k'} \delta_{n_{P(3)},l'}
\over E-E_{n_1}-E_{n_2}-E_{n_3}} \delta_{s_{P(1)},s_1'} 
\delta_{s_{P(2)},s_2'} \delta_{s_{P(3)},s_3'}.
\end{eqnarray}
Since these matrices are infinite dimensional ones we truncate them by
restricting $k,l$ between $-m$ and $m$, where $2m+1$ is the
number  of one-particle wave functions given by (\ref{fermion-one-partic}). 
Because of the spin indices each $(k,l),(k',l')$ elements of the
matrices M  still correspond to  8 by 8 matrices. Therefore 
in Eq. (\ref{polus-eq-fermion}) the hypermatrix is a 
$2 \cdot 8 \cdot (2m+1)^2$ dimensional matrix.
Grouping  these elements by the total  $S_z$ of the three fermions  
we may reduce the dimension of the matrix into a $ (2+3) \cdot
(2m+1)^2$ dimensional one. This dimensional reduction is possible
because we omit the energy independent factors in the matrix elements.  

The above given matrix elements still contain summations which can be
evaluated by using the following identity\cite{Abramowitz}:    
\begin{equation}
\sum_{n=-\infty}^{\infty} {1\over (n+\Phi)^2-z^2 }= {\pi\over 2 z}
\left(\cot \pi(\Phi-z)-\cot\pi(\Phi+z)\right).
\end{equation}

\begin{figure}[hbt]
{\centerline{\leavevmode \epsfxsize=8.5cm \epsffile{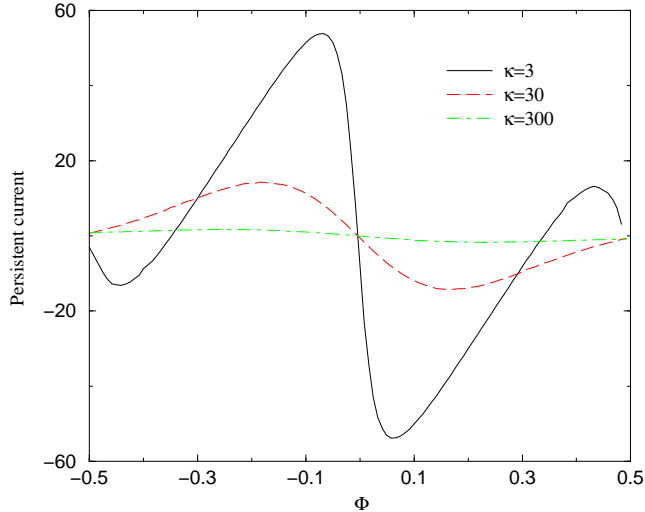 }}}
\caption{The persistent current calculated by the SD method is plotted
(in arbitrary units) as a function of the flux (in units of
$\Phi_0=h/e$) for fixed interaction strength $\lambda=2$ and different 
impurity strength $\kappa$. 
We used  $2m+1=5$ one-particle wave functions. 
\label{pc-abra}}
\end{figure}

In Fig.\ \ref{pc-abra} the persistent currents obtained by our SD
method as a function of flux in units of flux quantum $h/e$ 
are shown for different impurity strengths $\kappa$. In this
calculation $2m+1=5$ one-particle wave functions are used.
One can see that the persistent current suppresses with 
increasing the strength of the impurity. Our purpose in this paper is
to demonstrate how our SD method can be applied to different problems
such as the calculation of the persistent current. A more complete
analysis of the persistent current problem is beyond the scope of this
paper. The work along this line is in progress.

\section{Conclusions} \label{veg}

In this paper the Green's functions of  many-body systems 
were calculated when the interactions between the particles are Dirac-delta
type contact potentials, and when the system contains 
impurities described by contact potentials. 
The Hamiltonian of the system is taken as a sum of two parts, 
the first of which contains only the sum of the one-particle Hamiltonians, 
while the second contains the rest, namely the interactions
between the particles and the potentials of the impurities.   
Using the Dyson equation for the Green's function the iterative solution
for the Green's function can be formally given as the ratio of two 
determinants. However, in this formal expression the matrix elements of 
the determinants are `indexed' by continuous variables. 
To handle these determinants one
can use any complete orthogonal set of wave functions and the
corresponding operators in the determinants can be given in the usual 
matrix representation. 
This makes the numerical calculation of the Green's function possible. 

After presenting the general formalism for deriving the Green's function, 
several examples were discussed in order to show the relevant steps 
in the algebra as well as to test the method. 
Only the energy eigenvalues were calculated for different
systems. However, in our method the full Green's function is given, 
therefore, in principle, it can be used for calculating other types of 
physical quantities. The energy levels can be found by determining 
the poles of the Green's function. Since the Green's function of the 
system is given by the ratio of two determinants, one only needs to 
calculate the zeros of the spectral determinant (in the denominator of
the Green's function) to obtain the energy eigenvalues of the system.
 
For testing our method, we used a one-dimensional model of 
bosons interacting via Dirac-delta interactions. In this case 
the energy eigenvalues can be calculated exactly from the well known 
Bethe ansatz solution . In the case of two interacting bosons it was
shown algebraically that the SD method yields 
{\it exactly\/} the same energy levels as those obtained from the 
Bethe ansatz solution. For the case of three and four interacting bosons 
an equation for determining the energy levels was derived and then it was 
solved numerically. The results obtained from our method are in 
very good agreement with  those obtained from
the Bethe ansatz solution. 
A comparison of our method  with the direct diagonalization of 
the Hamiltonian was also presented. The same number of one-particle 
wave functions was used in the latter method as in the SD 
approach. It was demonstrated that our method gives, in general, ten times
smaller relative errors than the diagonalization method.
The energy of the ground state of the system can be obtained with even
better accuracy in spite of the fact that only a few one-particle 
wave functions were used (typically $9$; the relative errors were then 
less than $0.01\%$). As a nontrivial example we calculated the energy
levels of a system including two interacting bosons and one impurity. 
Both the interactions and the potential of the impurity were of Dirac-delta 
type. Applying our general method, we derived an equation for the
energy levels of the system. Since no Bethe ansatz type of solution
is known in this case, by increasing the number of one-particle wave
functions we calculated the energies by making the result convergent to 4
significant figures. Taking these energy levels as the exact ones, 
we compared our method with direct diagonalization. 
A much better accuracy was found in the SD method 
(e.g.\ for the ground state with $9$ one-particle wave functions 
the relative error was about $0.003\%$, while it was $2.6\%$ with the 
diagonalization method).     

Our method was also applied to spin-full fermion systems. 
A better accuracy was achieved in calculating the ground state
energy comparing the diagonalization method. Therefore our method 
is expected to be suited for numerical calculations for the persistent
current. We demonstrated the applicability of our SD method for the
case of three interacting fermions and one impurity, and calculated the
persistent current. We defer a thorough study of the persistent
current in a later publication. 
 
In summary, we developed a spectral determinant method with which much
more accurate values of the energy levels can be obtained numerically
than with the usual direct diagonalization method.
In our formalism the full Green's function
is available, thus it might be used in the TIP and persistent current 
problem mentioned in the introduction.

\acknowledgements 

This work was supported by the Hungarian Science Foundation
OTKA (TO25866 and TO34832), the Hungarian Ministry of Education and 
the TRN programme  HPRN-CT-2000-00144 of the EU. 
One of us (J.\ Cs.) thanks to A.\ Pir{\'o}th for his valuable comments 
on the manuscript. 

\appendix 
\section{Two interacting particles with $M$ impurities} \label{app-1}

In this section an expression for the GF is derived for the case of two 
interacting particles and $M$ impurities. 
The interaction Hamiltonian $H_1$ is given by 
\begin{equation}
H_1=\sum \limits_{i=1}^{M} \kappa_i \Bigl(\delta(x_1-u_i)+ \delta(x_2-u_i) 
\Bigr)+\lambda \ \delta(x_1-x_2), 
\end{equation}
where $u_i$ is the location of the $i$th impurity, of strength $\kappa_i$ 
and $\lambda$ is the strength of the interaction between
the particles. As in the case of a single impurity, the first-order
correction (in $H_1$) of the iterative solution of the Dyson equation is   
\begin{equation}
G^{(1)} = \int G_0(x_1,x_2|y_1,y_2) \ H_1(y_1,y_2) \ G_0(y_1,y_2|x_1',x_2') 
\ dy_1 dy_2.
\end{equation}
Inserting $H_1$ into the above expression the integrals can be
performed exactly for Dirac delta potentials.  Collecting the 
identical terms arising from permutational symmetry, the first order 
correction of the GF is 
\begin{equation}
G^{(1)} = \int \sum \limits_{i=1}^M 2 \kappa_i \ 
G_0(x_1,x_2|y,u_i)  G_0(y,u_i|x_1'x_2')  
+ \lambda \ G_0(x_1,x_2|y,y)  G_0(y,y|x_1',x_2') \ dy.
\end{equation}
Similarly, the second order correction takes the form
\begin{equation}
G^{(2)} = \int G_0(x_1,x_2|y_1,y_2) \ H_1(y_1,y_2) \ G_0(y_1,y_2|y_1',y_2') 
\ H_1(y_1',y_2') \ G_0(y_1',y_2'|x_1',x_2') \ dy_1 dy_2 dy_1' dy_2'.
\end{equation}
Higher order corrections can be easily found. 
In a way similar to the case of $G^{(1)}$ we have
\begin{eqnarray}
G^{(2)} &=& \int \sum \limits_{i,j=1}^M 2 \kappa_i \ G_0(x_1,x_2|y,u_i) 
 2\kappa_j \ G_0(y,u_i|y',u_j) G_0(y',u_j|x_1'x_2') 
\ dy dy'   \nonumber \\
&+& \int \sum \limits_{i=1}^M 2 \kappa_i \ G_0(x_1,x_2|y,u_i)  
 \lambda \ G_0(y,u_i|y',y') G_0(y',y'|x_1'x_2') \ dy dy'  
\nonumber \\
&+& \int \sum \limits_{j=1}^M \lambda \ G_0(x_1,x_2|y,y)  
 2\kappa_j \ G_0(y,y|y',u_j) G_0(y',u_j|x_1'x_2') \ dy dy' 
 \nonumber \\
&+& \int \lambda \ G_0(x_1,x_2|y,y)   \lambda \ 
G_0(y,y|y',y') G_0(y',y'|x_1'x_2') \ dy dy'.  
\end{eqnarray}
It can be shown that the GF including the higher order corrections is 
\begin{eqnarray}
G &=& G_0 + \int \sum_{\alpha=1}^{M+1} A_\alpha (y) \ B_\alpha (y) \ dy+  
\int \sum_{\alpha,\beta=1}^{M+1} A_\alpha (y) \ C_{\alpha\beta} (y,y') 
\ B_\beta (y') \ dy dy' + \dots  \nonumber \\ 
&=& G_0 + \int \sum_{n=0}^\infty \sum_{\alpha,\beta=1}^{M+1} A_\alpha (y) \ 
\Bigl(C_{\alpha\beta} (y,y')\Bigr)^n \ B_\beta (y') \ dy dy',
\end{eqnarray}
where 
\begin{equation}
A_\alpha (y) = \left( \matrix{ 2 \kappa_i \ G_0(x_1,x_2|y,u_i) \cr \lambda 
\ G_0(x_1,x_2|y,y)} \right),\ \  B_\beta (y') = \left( 
\matrix{ G_0(y',u_j|x_1',x_2') \cr G_0(y',y'|x_1',x_2')} \right),
\label{A-B}
\end{equation}
are $(M+1)$-component vectors ($i=1,\dots,M$) and
\begin{equation}
C_{\alpha\beta} (y,y') =  \left( \matrix{ 2 \kappa_j \ G_0(y,u_i|y',u_j) & 
\lambda \ G_0(y,u_i|y',y') \cr 2 \kappa_j \ G_0(y,y|y',u_j) &  \lambda \ 
G_0(y,y|y',y')} \right),
\label{C-matrix} 
\end{equation}
is an $(M+1) \times (M+1)$ matrix. 
This is a Neumann series which can be summed:
\begin{equation}
G = G_0+\int \sum_{\alpha,\beta=1}^{M+1}  A_\alpha (y) \ \Bigl(\delta(y-y') 
\delta_{\alpha\beta}-C_{\alpha\beta} (y,y')\Bigr)^{-1} \ B_\beta (y') 
\ dy dy'.
\end{equation}
Using  Eqs.\ (\ref{A-B}), (\ref{C-matrix}), and the matrix identity in 
Eq.\ (\ref{hyperm}), the GF can be rewritten 
as the ratio of two determinants:
\begin{equation}
G(x_1,x_2|x_1',x_2') ={\det \left[ \matrix{G_0(x_1,x_2|x_1',x_2') & 
G_0(x_1,x_2|y',y') & G_0(x_1,x_2|u_1,y')& \cdots & G_0(x_1,x_2|u_M,y')
 \cr
G_0(y,y|x_1',x_2') & L(y|y') & G_0(y,y|u_1,y') & \cdots & G_0(y,y|u_M,y')
\cr 
G_0(u_1,y|x_1',x_2') & G_0(u_1,y|y', y') & K_1(y|y') & \cdots 
& G_0(u_1,y|u_M,y') \cr \vdots & \vdots & \vdots & \ddots & \vdots \cr
G_0(u_M,y|x_1',x_2') & G_0(u_M,y|y',y') & G_0(u_M,y|u_1,y') & 
\cdots  & K_M(y|y')}
 \right] \over \det \left[ \matrix{ L(y|y') &
G_0(y,y|u_1, y') & \cdots   & G_0(y,y|u_M, y') \cr G_0(u_1, y|y',y')
 & K_1(y|y') & \cdots  & G_0(u_1,y|u_M,y') \cr \vdots & \vdots & 
\ddots & \vdots \cr
G_0(u_M, y|y',y') & G_0(u_M, y|u_1, y') &\cdots & K_M(y|y')  } \right] }, 
\end{equation}
where
\begin{equation}
K_i(y|y') = -{1 \over 2 \kappa_i} \ \delta (y-y') +  G_0(u_i, y|u_i',y'),
\end{equation}
and
\begin{equation}
L(y|y) = -{1 \over \lambda} \delta(y-y') + G_0(y,y|y',y'). 
\end{equation}

\section{Two interacting bosons in one dimension} \label{2boson}
In this subsection an equation for the energy levels of a system
with two interacting electrons in one dimension with periodic boundary
conditions is derived. It turns out that this equation is the same 
as that obtained from the Bethe ansatz solution. Thus, the two 
methods lead to the same energy levels in this case. 

Consider a one dimensional box of length $a$, with periodic boundary 
conditions. 
The one-particle Hamiltonian is given by Eq.\ (\ref{free-H}), the
potential $V(x)$ is zero. Using periodic boundary conditions the 
one-particle eigenfunctions of this Hamiltonian are given by
\begin{equation}
\psi_k(x)=   {1\over \sqrt a} \ 
e^{2 \pi i k x/a},
\label{1wave}
\end{equation}
and the eigenvalues are 
\begin{equation}
E_k=  \frac{4\pi^2k^2}{a^2},
\label{E_n}
\end{equation}
where $k=0,\pm 1,\pm 2, \cdots$.
From these wave functions one can construct the bosonic two-particle wave
functions which may be written as
\begin{equation}
\psi_{n,m}(x,y)= \left\{ \begin{array}{ll}
\frac{\psi_n(x) \psi_m(y)+ \psi_n(y)\psi_m(x)}{\sqrt 2},  
& \mbox{for }\,\,\,\, n > m,  \\[1ex]
\psi_n(x) \psi_m(y),& \mbox{for} \,\,\,\, n=m.
\end{array}  \right.  
\label{2wave}
\end{equation}
For two interacting particles the GF and the operator $L(y_1,y_2)$ are given 
in Eqs.\  (\ref{teljes-GF}) and (\ref{L-alak}), respectively.
Using Eqs.\ (\ref{2p-G0}) and (\ref{2wave}) the  unperturbed two-particle 
GF and the operator $L(y_1,y_2)$ can be expressed in terms of the one
particle eigenfunctions given in Eq.\ (\ref{1wave}): 
\begin{eqnarray}
G_0(x,x'|y,y') & = & 
\frac{1}{2} \sum_{n,m} \frac{\psi_n(x) \psi_m(x') 
\psi_n^*(y) \psi_m^*(y')}{E-E_n-E_m } + 
\frac{\psi_n(x) \psi_m(x') \psi_m^*(y) \psi_n^*(y')} 
{E-E_n-E_m }, \\
L(y_1,y_2) & = & -\frac{1}{\lambda}\, \,  \delta (y_1-y_2)+ 
\sum_{n,m} 
\frac{\psi_n(y_1) \psi_m(y_1) \psi_n^*(y_2) \psi_m^*(y_2)}{ E-E_n-E_m },
\label{L2} 
\end{eqnarray}
where the strength of the interaction between the two particles is 
$\lambda$, and $E_n$ is given in Eq.\ (\ref{E_n}).
 
The  poles of the Green's function given in Eq.\ (\ref{teljes-GF}) 
are the energy eigenvalues of 
the system. The equation determining the
poles of GF  is 
\begin{equation}
\det L(y_1|y_2) = 0,
\label{det2L}
\end{equation}
where the operator $L(y_1|y_2)$ is given by Eq.\ (\ref{L2}).
To evaluate the determinant it is convenient to use the basis given 
by Eq.\ (\ref{1wave}). The matrix element $L_{kl}$ of the operator 
$L(y_1|y_2)$ is then 
\begin{equation}
 L_{kl} = \int_0^a \ 
\psi_k^*(y_1) \ L(y_1,y_2) \ \psi_l(y_2)\ dy_1 \ dy_2 \ = 
 \delta_{k,l} 
\left(- {1\over \lambda}+ {1 \over a}\sum_n {1\over E-E_n-E_{k-n}} 
\right),
\end{equation} 
where we have used the integral
\begin{equation}
\int_0^a \psi^*_k(x) \ \psi_n(x) \ \psi_m(x) \ dx = 
{1 \over {\sqrt a}} \  \delta_{n+m-k,0}.
\end{equation}
Since the matrix $L_{kl}$ is diagonal, its determinant is 
\begin{equation}
\det L_{kl} = \prod_{k= - \infty}^{+\infty} 
\left(- {1\over \lambda}+ {1 \over a} \sum_{n= -\infty}^{+\infty} 
{1\over E-E_n-E_{k-n}} \right).
\end{equation}
Thus, Eq.\ (\ref{det2L}) is equivalent to 
\begin{equation}
{a \over \lambda} = \sum_{n=-\infty}^{+\infty} {1 \over E- 
{4\pi^2 \over a^2}\left(n^2+(k-n)^2 \right)}, 
\label{eq-2p}
\end{equation}
where $k=0, \pm 1,\pm 2, \dots $.
Using the following identities\cite{Abramowitz} 
\begin{eqnarray}
\sum_{n=-\infty}^{+\infty} {1\over z^2-(2n)^2} & = & 
{\pi \over 2z} \cot {\pi \over 2} z, 
\nonumber \\
\sum_{n=-\infty}^{+\infty} {1\over z^2-(2n+1)^2} & = & 
-{\pi \over 2z} \tan {\pi \over 2} z.
\label{azonossag}
\end{eqnarray}
the right hand side of Eq.\ (\ref{eq-2p}) can be
written as 
\begin{eqnarray}
2\, \frac{\sqrt{2 E - 4\pi^2 k^2/a^2}}{\lambda} & = & 
\cot {a \over 4}\sqrt{2 E - 4\pi^2 k^2/a^2}
\,\,\,\, \mbox{ for $k$ even,}  
\label{1sol} \\
2\, \frac{\sqrt{2 E - 4\pi^2 k^2/a^2}}{\lambda} & = &
-\tan {a \over 4} 
\sqrt{2 E - 4\pi^2 k^2/a^2} 
\,\,\,\, \mbox{ for $k$ odd.}
\label{2sol}
\end{eqnarray}

This result is equivalent to the Bethe ansatz solution
\cite{Lieb-I,Mattis} for a one-dimensional Bose gas with Dirac delta 
interactions.
According to the Bethe ansatz solution, the energy eigenvalues $E$ of 
the $N$ interacting particles can be determined 
from the solution for $k_j$ of the following $N$ equations:  
\begin{equation}
{(-1)}^{N-1} e^{-i k_j a} = \exp \left( i \sum_{s=1}^N -2 \arctan {k_s-k_j 
\over \lambda/2 } \right), 
\label{Bethe-eq}
\end{equation}
where $j=1,2,\dots,N$ and  
\begin{equation}
E = \sum \limits_{j=1}^N k_j^2.
\label{Bethe-E}
\end{equation}

For two particles, the two equations for $k_1$ and $k_2$ can be rewritten as
\begin{eqnarray}
e^{-i P a} & = & 1, \\
e^{i a \sqrt {2E-P^2}} & = & \exp \left( -4 i 
\arctan{\sqrt{2E-P^2}\over \lambda/2} \right),
\end{eqnarray}
where $P= k_1+k_2$ is the total momentum of the particles. 
From the first equation we have $P=2\pi n/a$, where $n$ is an integer, and
the second equation may be written as
$a \sqrt{ 2E- P^2}+ 2 \pi n_1 
= - 4 \arctan {\sqrt {2 E-P^2} \over \lambda/2 }$,
where $n_1$ is another integer.
Taking the tangent of this equation we find the same forms as given 
in Eqs.\ (\ref{1sol})-(\ref{2sol}) for even and odd $n_1$, respectively.


\begin{thebibliography}{10}

\bibitem{Dorokhov}
O.~N. Dorokhov, {Z}h.\ {\'E}ksp.\ Teor.\ Fiz. {\bf 98},  646  (1990), [Sov.
  Phys. JETP {\bf 71,} 360 (1990)].

\bibitem{Shepelyansky}
D.~L. Shepelyansky, Phys. Rev. Lett. {\bf 73},  2607  (1994).

\bibitem{Buttiker1}
M. B\"uttiker, Y. Imry, and R. Landauer, Phys. Lett. {\bf 96A},  365  (1983).

\bibitem{Levy}
L.~P. Levy, G. Dolan, J. Dunsmuir, and H. Bouchiat, Phys. Rev. Lett. {\bf 64},
  2074  (1990).

\bibitem{Chandrasekhar}
V. Chandrasekhar {\it et~al.}, Phys. Rev. Lett. {\bf 67},  3578  (1991).

\bibitem{Oppen1}
F. von Oppen, T. Wetting, and J. M\"uller, Phys. Rev. Lett. {\bf 76},  491
  (1996).

\bibitem{Song1}
P.~H. Song and D. Kim, Phys. Rev. B {\bf 56},  12217  (1997).

\bibitem{Ortuno}
M. Ortu$\tilde{\rm n}$o and E. Cuevas, Europhys. Lett. {\bf 46},  224
(1999).

\bibitem{Pichard1}
D. Weinmann, A. M\"uller-Groeling, J.-L. Pichard, and K. Frahm, Phys. Rev.
  Lett. {\bf 75},  1598  (1995).

\bibitem{Berkovits1}
R. Berkovits and Y. Avishai, Europhys. Lett. {\bf 29},  475  (1995).

\bibitem{Berkovits2}
D. Yellin and R. Berkovits, Phys. Rev. B {\bf 51},  4369  (1995).

\bibitem{Leadbeater}
M. Leadbeater, R.~A. R{\"o}mer, and M. Schreiber, Eur. Phys. J.  B  
{\bf 8}, 643 (1999). 

\bibitem{Chandross}
M. Chandross and J. C. Hicks, Phys. Rev. B {\bf 59},  9699  (1999).

\bibitem{Eckle}
H.-P. Eckle, A. Punnoose and R. A. R{\"o}mer, Europhys. Lett. {\bf 39},
293 (1997).


\bibitem{Grosche}
C. Grosche, Ann. Phys. (Leipzig) {\bf 2},  557  (1993).

\bibitem{Chaos-cikk}
G. Vattay, J. Cserti, G. Palla, and G. Sz\'alka, Chaos, Solitons \& Fractals
  {\bf 8},  1031  (1997).

\bibitem{Crossover}
J. Cserti, G. Sz\'alka, and G. Vattay, Phys. Rev. B {\bf 57},  R15092  (1998).


\bibitem{Bethe}
H.~A. Bethe, Z. Physik {\bf 71},  205  (1931).

\bibitem{Lieb-I}
E.~H. Lieb and W. Liniger, Phys. Rew. {\bf 130},  1605  (1963).

\bibitem{Lieb-II}
E.~H. Lieb, Phys. Rev. {\bf 130},  1616  (1963).

\bibitem{Yang-Yang}
C.~N. Yang and C.~P. Yang, J. Math. Phys. {\bf 10},  1115  (1969).

\bibitem{Mattis}
D.~C. Mattis,  in {\em The Many-Body Problem}, edited by D. C. Mattis (World
  Scientific Publishing Co.\ Pte., Singapore, 1993).

\bibitem{Economou}
E.~N. Economou, {\em {G}reen's functions in Quantum Physics}, 2nd ed.
  (Springer-Verlag, Berlin, Germany, 1983).

\bibitem{Arfken}
G.~B. Arfken and H.~J. Weber, {\em Mathematical Methods for Physicists}, 4th
  ed. (Academic Press, San Diego, CA, 1995).

\bibitem{Abramowitz}
M. Abramowitz and I. Stegun, {\em Handbook of Mathematical Functions}, 9th ed.
  (Dover Publication Inc., New York, NY, 1972), p.\ 228.

\bibitem{Cheung}
H. Cheung, Y. Gefen, E.~R. Riedel, and W. Shih, Phys. Rev. {\bf 37},  6050
  (1988).

\end{thebibliography}

\end{document}